# The Effect of Ionic Spin on Multiferroic of Orthorhombic Perovskite

Kaiyang Gao[1], Jiyu Shen[1], Zeyi Lu[1], Jiajun Mo[1], Guoqing Liu[1], Zhongjin Wu[1], Chenying Gong[1], Dong Xie[1], Yanfang Xia[1]* and Min Liu[1,2]*

[1]College of Nuclear Science and Technology, University of South China, Hengyang 421200, Hunan, P.R.China.

[2]Zhuhai Tsinghua University Research Institute Innovation Center, 101 University Ave, Tangjiawan Zhuhai 519000, Guangdong, P.R.China.

Email: xiayfusc@126.com, liuhart@126.com.

**Abstract**

To investigate the influence of ion spin on the coupling between ferromagnetism and ferroelectricity in type II multiferroic perovskite, we prepared the multiferroic perovskite $Er_{0.9}La_{0.1}Cr_{0.8}Fe_{0.2}O_3$ (ELCFO) using the sol-gel method, and explored the macroscopic magnetic properties of ELCFO through Mössbauer spectrum and magnetic testing. The thermal magnetic curve was analyzed to examine the state and change of each ionic spin in the ELCFO system at different temperature ranges, and the role of ionic spin in the coupling between ferromagnetism and ferroelectricity was investigated. This study provides a theoretical basis for further research on multiferroic perovskites and has practical implications.

**Key:** Perovskite, Spin, Dzyaloshinksii-Moriya Interaction, Magnetization, Polarization.

## 1.Introduction

Perovskite is an oxide with the structural formula of $ABO_3$[1], as illustrated in **Figure 1**. The A site can be a rare earth ion (e.g., Pr, Dy, Er, etc.), while the B site can be a transition metal ion (e.g., Fe, Cr, Mn, etc.). Due to their excellent physical properties, such as photosensitivity, electrical and magnetic properties, perovskites have been extensively studied in various fields[2-5]. Previous studies have shown that the physical properties of pristine perovskites can be tailored by doping with new ions, such as modifying their structure[6], inducing ferroelectricity[7,8], or introducing ferromagnetism[9, 10].

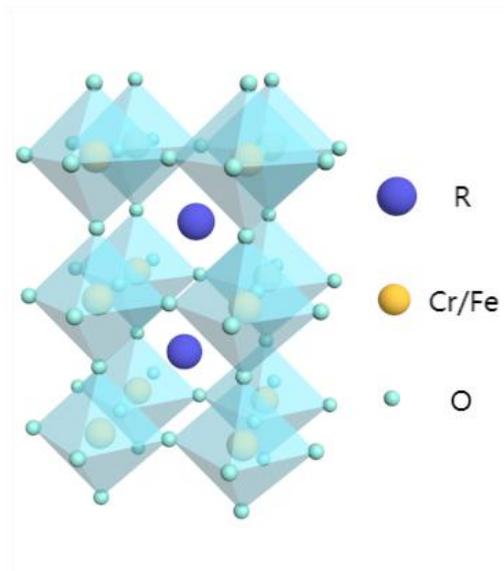

**Fig. 1** Schematic diagram of perovskite structure.

While many studies have investigated the ferroelectricity or ferromagnetism of perovskites, achieving the coexistence of these two properties remains a challenge due to their conflicting mechanisms[11,12]. Ferroelectricity arises from the absence of electrons in the 3d shell, which favors hybridization of the electron cloud of adjacent ions, whereas ferromagnetism results from the appearance of a magnetic moment by arranging some electrons in the 3d shell layer of the transition metal. As a result of these opposing mechanisms, most materials can only exhibit either ferroelectricity or ferromagnetism, with the exception of multiferroic materials (classified as Type I and Type II), which can simultaneously exhibit both properties[13-15]. Type II multiferroic materials, in particular, have a solid magnetoelectric coupling due to their ferroelectric polarization originating from a unique magnetic order[16]. This property is of great importance for developing new types of information memory[17].

Type II multiferroic perovskite exhibits a paraelectric-paramagnetic state before undergoing a phase transition. Ferroelectricity and ferromagnetism coexist when the temperature is lowered below the phase transition temperature point $T_N$[18-21]. After the phase transition, most of the perovskite shows antiferromagnetic order due to the super-exchange effect, in which the magnetic moments of adjacent transition metals are reversed and parallel. However, the Dzyaloshinksii-Moriya Interaction (DMI) causes adjacent magnetic moments to be perpendicular, resulting in a weaker ferromagnetic order in the perpendicular direction of the antiferromagnetic order, which leads to weak ferromagnetism in the system[22-23]. DMI also generates partial polarization: the DMI of adjacent magnetic moments causes displacement of the oxygen ions between them with respect to the cations to reduce their energy, which leads to displacement polarization. In addition, lattice distortion causes rare earth ions within the sublattice to be closest to the transition metal ions on the adjacent side. Due to the antiferromagnetic arrangement according to the G-type, the spins between adjacent transition metal ions are in opposite directions, allowing the rare earth ions to form a stable polarization. The rare earth ion couples with its closest transition metal ion, and if the spin of that transition metal ion changes, the spin of the rare earth ion also changes.

In this paper, we prepared a sample of $Er_{0.9}La_{0.1}Cr_{0.8}Fe_{0.2}O_3$ perovskite with different spin structures by the sol-gel method, doping $ErCrO_3$ with a small amount of La and Fe[24-25]. We explored the physical characteristics of the samples by XRD, Mössbauer spectrum, and magnetic testing. Based on the variation of the thermomagnetic curves, we analyzed the spin states of ions in the ELCFO system at different temperatures and used this to explore the origin of magnetization and polarization in this system.

## 2.Experiment procedure

### 2.1. Making samples

We prepared $Er_{0.9}La_{0.1}Cr_{0.8}Fe_{0.2}O_3$ nanopowders using a sol-gel method. Lanthanum nitrate hexahydrate, erbium nitrate hexahydrate, chromium nitrate nonahydrate, and iron nitrate nonahydrate were dissolved in deionized water according to the stoichiometric ratio, and citric acid and ethylene glycol were added in deionized water at a ratio of metal ion:citric acid:ethylene glycol = 1:1.2:1.2. The resulting sol was thoroughly stirred at 80°C, dried at 150°C for 4 hours, and finally calcined at 1200°C for 12 hours. The $Er_{0.9}La_{0.1}Cr_{0.8}Fe_{0.2}O_3$ nanopowder was obtained after complete grinding.

*2.2 Sample characterization*

*2.2.1. X-ray diffraction measurements*

X-ray diffraction measurements(XRD) in Siemens D500 Cu Kα($\lambda$=1.54178Å). Diffraction angle range from 20° to 80°(every 0.5s forward step, each step forward 0.02°). The obtained data were refined using FullProf software.

*2.2.2. Mössbauer spectrum test*

The transmission $^{57}Fe$ Mössbauer spectrum of $Er_{0.9}La_{0.1}Cr_{0.8}Fe_{0.2}O_3$ were collected at RT on SEE Co W304 Mössbauer spectrometer with a $^{57}Co/Rh$ source in transmission geometry equipped in a cryostat (Advanced Research Systems, Inc., 4 K). The data results were fitted with MössWinn 4.0 software. The obtained data were refined using FullProf software.

*2.2.3. Magnetic test*

Magnetization measurements were done using a vibrating sample magnetometer (VSM) from Quantum DesignTM. The magnetization curves (M-H) are obtained at temperatures 300 K, 50K and 5K. The thermomagnetic curves (M-T) of the samples $Er_{0.9}La_{0.1}Cr_{0.8}Fe_{0.2}O_3$ were obtained under an external magnetic field of 100 Oe. Field-cooled (FC) and zero-field-cooled (ZFC) measurements were done at temperatures ranging from 4 K to 300 K.

**3.Results and Discussions**

## 3.1. Structural Analysis

The refined XRD patterns of $Er_{0.9}La_{0.1}Cr_{0.8}Fe_{0.2}O_3$ are presented in **Figure 2**. The analysis reveals that ELCFO belongs to the orthorhombic crystal system with a space group of Pbnm and lattice parameters $a$ = 5.257(3)Å, $b$ = 5.530(9)Å, $c$ = 7.563(9)Å, and $V$ = 219.938Å$^3$. The diffraction peak of ELCFO is shifted to the left compared to ECO[19]. This is attributed to the larger radii of the doped $La^{3+}$(1.36Å) and $Fe^{3+}$(0.645Å) ions, in comparison to those of $Er^{3+}$(0.89Å) and $Cr^{3+}$(0.615Å) ions[26]. The valence states and the radii of the ions in the ELCFO are presented in **Figure 3** and **Table 1**.

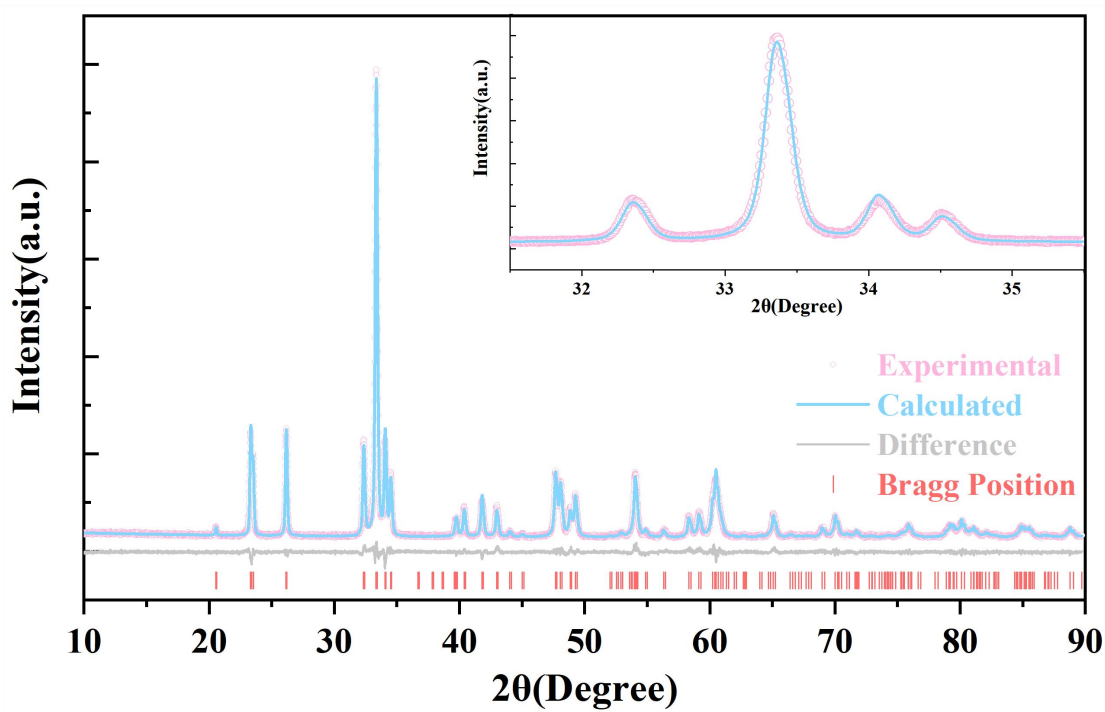

**Fig. 2** Rietveld refinement plot of $Er_{0.9}La_{0.1}Cr_{0.8}Fe_{0.2}O_3$.

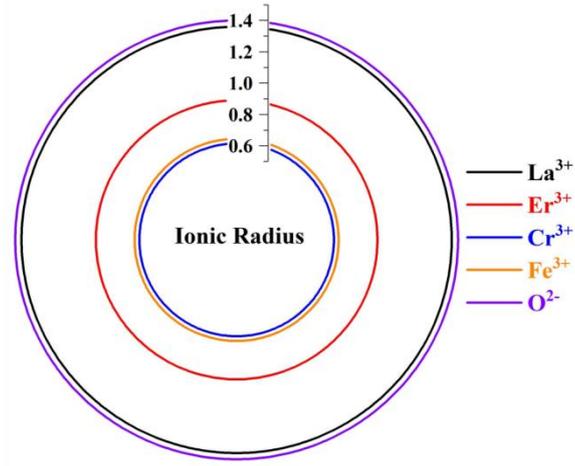

**Fig. 3** Schematic Diagram of Ion Radius

**Table. 1** Ion radius, valence state and coordination number.

| Ion | Coordination | Charge | Ionic Radius |
|-----|--------------|--------|--------------|
| La  | 12           | 3+     | 1.36         |
| Er  | 12           | 3+     | 0.89         |
| Cr  | 6            | 3+     | 0.615        |
| Fe  | 6            | 3+     | 0.645        |
| O   | 6            | 2-     | 1.4          |

The distortion of ELCFO can be evaluated using the Goldschmidt tolerance factor[27]:

$$\tau = (0.9R_{Er} + 0.1R_{La} + R_O) / \sqrt{2}(0.8R_{Cr} + 0.2R_{Fe} + R_O) \quad (1)$$

Upon calculation, we obtain $\tau$=0.817. This value suggests that the perovskite undergoes a significant degree of lattice distortion, which is responsible for the orthorhombic structure of the ELCFO system.

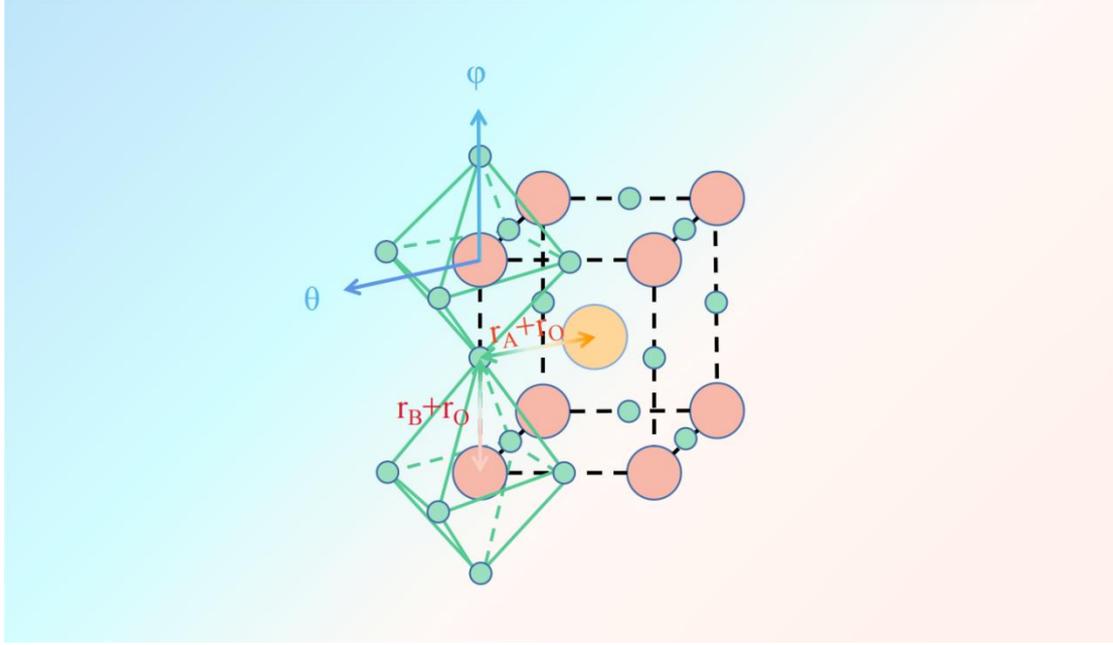

**Fig. 4** The distance between A-O ions and B-O ions and the torsion angle of the direction axis relative to [110] and [001] $\theta$ and $\varphi$.

At the same time, the octahedron also rotates. To describe the rotation of the octahedron, we construct two coordinate axes along the [110] direction and the [001] direction (as shown in **Figure 4**) to represent the rotation angle of the octahedron along the axis[28]:

$$\theta = cos^{-1}(a/b) \quad (2)$$

$$\varphi = cos^{-1}(\sqrt{2}\,a/c) \quad (3)$$

Upon calculation, we obtain $\theta$=18.097° and $\varphi$=10.595°, it shows that the octahedron rotates. Due to the lattice distortion, there are microstresses between crystals. The microstress $\varepsilon$ between the crystals and the average crystal size $D$ can be calculated using the Williamson-Hall (WH) formula:

$$cos\theta \beta_T = \varepsilon(4sin\theta) + \lambda K/D \quad (4)$$

Where $\lambda$ represents the wavelength of X-rays ($\lambda$ = 1.54178Å), $K$ = 0.89, $\beta_T$ is the half-height width at the peak, and *θ* is the diffraction angle. **Figure 5** shows the W-H image, and based on its slope and intercept, we determined the microstress to be $\varepsilon$ = 0.00233(4) and the lattice spacing to be ***D*** = 65.97(1) nm.

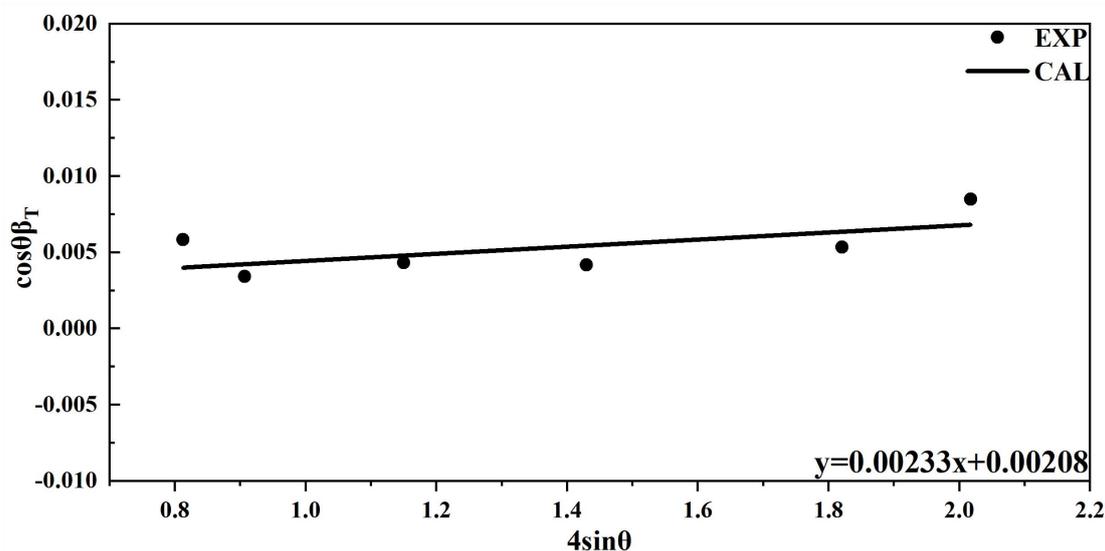

**Fig. 5** Williamson–Hall analysis of $Er_{0.9}La_{0.1}Cr_{0.8}Fe_{0.2}O_3$.

*3.2. Mössbauer spectrum*

At room temperature, the Mössbauer spectrum image of $Er_{0.9}La_{0.1}Cr_{0.8}Fe_{0.2}O_3$ is presented in **Figure 6**. The spectrum exhibits a double peak, which characterizes the paramagnetic state of the sample. The hyperfine parameters IS and QS were obtained through fitting. The value of IS = 0.228 suggests that all Fe ions are $Fe^{3+}$ in a high spin state. The presence of QS = 0.265 indicates an electric field gradient around $Fe^{3+}$, which results from the distortion and rotation of the octahedron in which $Fe^{3+}$ is located.

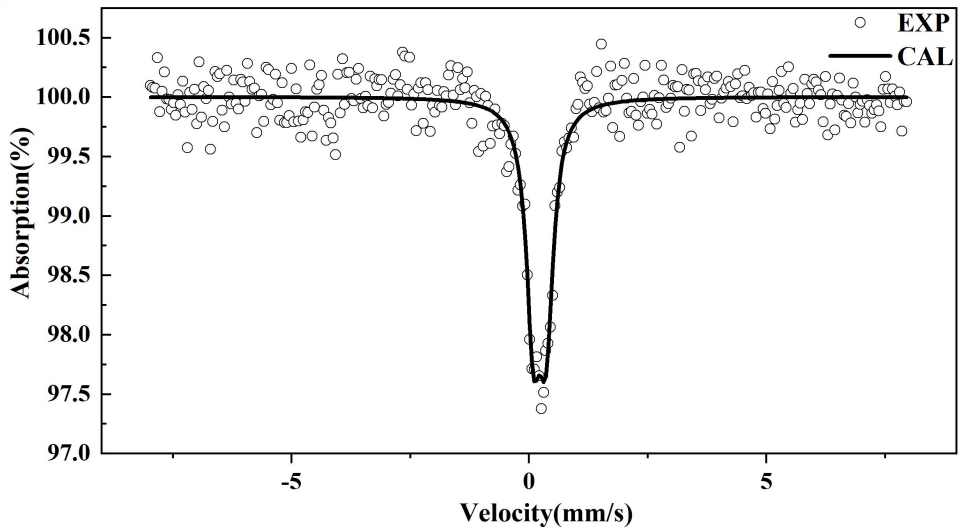

**Fig. 6** The Mössbauer spectrum image at room temperature.

### *3.3. Magnetic test*

**Figure 7** illustrates the magnetization curves of ELCFO at different temperatures ($T$ = 5K, 50K, 300K). At $T$ = 300K, the magnetization intensity shows a linear relationship with the magnetic field, indicating a paramagnetic state, which is consistent with the results from Mössbauer spectroscopy. At $T$ = 50K, a small hysteresis loop appears in the magnetization curve, indicating the presence of ferromagnetic order induced by weak ferromagnetic component between $Cr^{3+}/Fe^{3+}$ due to Dzyaloshinskii-Moriya interaction (DMI)[29]. At $T$ = 5K, a typical S-shaped magnetization curve with hysteresis loop is observed, which has also been observed in other multiferroic materials[30, 31]. Due to the larger magnetic moment of $Er^{3+}$ compared to $Cr^{3+}/Fe^{3+}$, the difference in antiparallel magnetic moments increases with decreasing temperature. Therefore, the magnetization increases with decreasing temperature[32, 33].

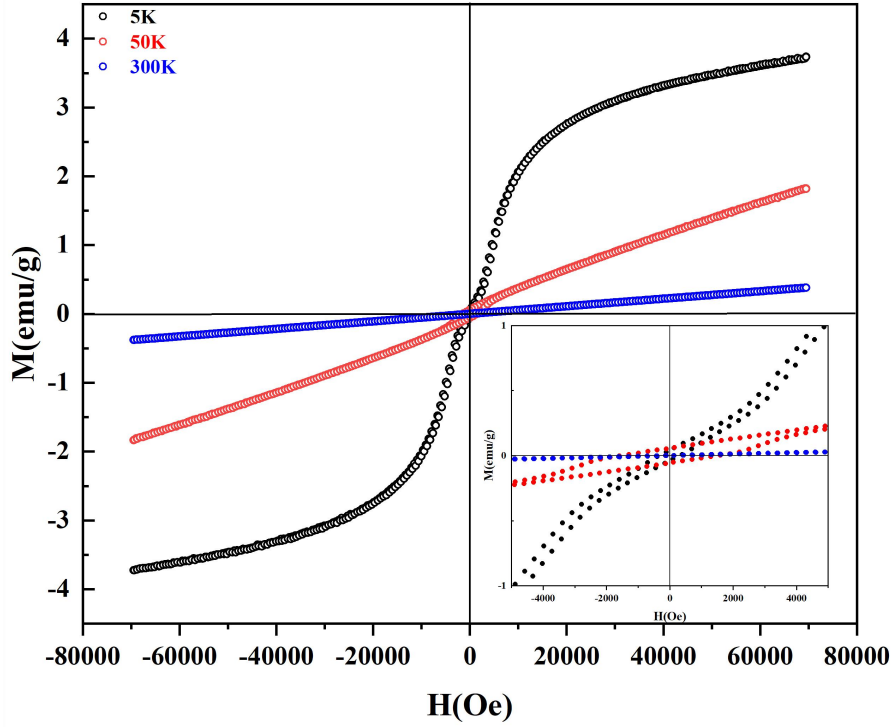

**Fig. 7** Magnetization curves of $Er_{0.9}La_{0.1}Cr_{0.8}Fe_{0.2}O_3$ at 300K, 50K and 5K.

**Figure 8** shows the variation of ELCFO magnetization intensity from 4 to 300K under an applied magnetic field of 100Oe. Two magnetic transitions can be observed clearly at $T_N$ ~ 147K and $T_{SR}$ ~ 13K, where $T_N$ is the Néel temperature of the sample from paramagnetic to antiferromagnetic state, and $T_{SR}$ is the reported spin-reorientation point where magnetic rare-earth ion $Er^{3+}$ couples with $Cr^{3+}/Fe^{3+}$[34-38]. Compared with the thermal magnetization curve of $ErCrO_3$ ($T_N$ ~ 133K), $T_N$ of $Er_{0.9}La_{0.1}Cr_{0.8}Fe_{0.2}O_3$ appears at a higher temperature (147K), which is due to the fact that the Néel temperature of Fe ($T_N^{Fe}$ ~ 625K) is higher than that of Cr ($T_N^{Cr}$ ~ 130K).

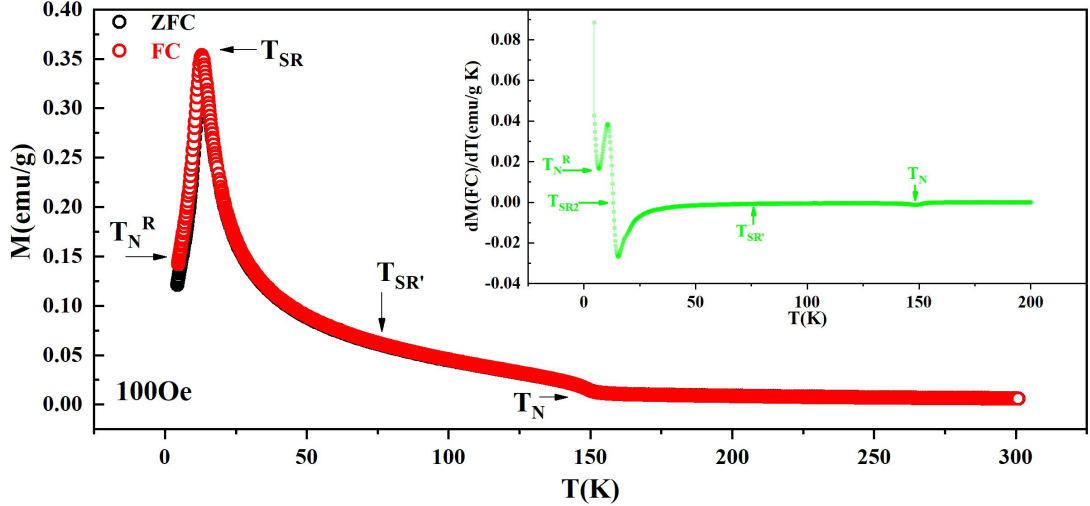

**Fig. 8** Thermomagnetic curves of $Er_{0.9}La_{0.1}Cr_{0.8}Fe_{0.2}O_3$ under $H$=100Oe. The illustration shows *dM/dT* vs. *T*.

ELCFO exhibits complex magnetic behavior within the testing range. The FC/ZFC curve shows an increasing trend with decreasing temperature, reaching a maximum around $T_{SR}$ ~ 13K before starting to decrease. Above $T_N$, ELCFO is paramagnetic, consistent with the room-temperature Mössbauer spectrum. Below $T_N$, $Cr^{3+}$ and $Fe^{3+}$ align along the z-axis according to $\Gamma_4$, providing a weak ferromagnetic component. The magnetic moment component of $Er^{3+}$, a magnetic rare-earth ion, also contributes to the magnetization strength along the weak ferromagnetic direction. As the temperature decreases, the B-site ions undergo spin reorientation, and the spin structure transitions from $\Gamma_4$ to $\Gamma_2[\Gamma_4(G_x, A_y, F_z) \to \Gamma_{24}(G_x, G_z) \to \Gamma_2(F_x, C_y, G_z)]$. At $T_{SR'}$, the B-site ions complete their spin reorientation and align according to the $\Gamma_2$ symmetry[39 40]. Previous studies reported the presence of $\Gamma_2$ allowed in $ErCrO_3$[41] and observed the transition from $\Gamma_4$ to $\Gamma_2$ in $ErFeO_3$ and $ErCr_{0.5}Fe_{0.5}O_3$ as well[42 43]. As the temperature drops below $T_{SR}$, the B-site ions align according to the $\Gamma_{12}(G_y, G_z)$ symmetry, and the magnetization rapidly decreases[43]. This change in spin direction is caused by the Dzyaloshinskii-Moriya interaction(DMI) and the anisotropic magnetic interactions between $Er^{3+}$ and $Cr^{3+}/Fe^{3+}$[44-46]. Since $La^{3+}$ is a non-magnetic ion, doping does not change the type of spin reorientation but can adjust the temperature range of this phase transition[47].

By observing the inset of Figure 8, it can be noticed that there are significant changes in *dM/dT* at $T_N$ and $T_{SR}$. The inset shows how the magnetization changes in the region between $T_{SR}$ and $T_{SR'}$ as $Cr^{3+}/Fe^{3+}$ undergo spin-reorientation. At *T*~4.9K, the *dM/dT* vs. *T* curve has a minimum value. At this temperature, $Er^{3+}$ becomes antiferromagnetically ordered, labeled as $T_N^R$[48]. This spin reordering is caused by the $Er^{3+}$-$Er^{3+}$ superexchange interaction[49], which is difficult to directly observe in the thermal magnetic curve but cannot be ignored.

### *3.4. Multiferroic origin*

The thermal-magnetic curve shows that ELCFO undergoes a phase transition at $T_N$, and the emergence of magnetization and polarization has been observed[19]. This transition indicates that ELCFO changes from a paraelectric-paramagnetic state to a ferroelectric-antiferromagnetic state. After the phase transition, ELCFO arranges into a G-type antiferromagnetic order, where adjacent $Cr^{3+}/Fe^{3+}$ magnetic moments are inverted parallel aligned, and the spin direction is arranged according to $\Gamma_4$. Due to the difference in magnetic moments between $Fe^{3+}$(5.92 μB) and $Cr^{3+}$(3.87 μB), magnetization exists in the direction parallel to the x-axis. In addition, the magnetic moment of $Er^{3+}$ contributes to the magnetization of ELCFO. Due to the existence of DMI, adjacent magnetic moments are tilted, and the $Cr^{3+}/Fe^{3+}$ magnetic moments arranged according to the $\Gamma_4$ spin structure will generate a ferromagnetic component along the z-axis direction. At the same time, the $O^{2-}$ ions between the cationic connection lines will experience a displacement relative to the cation, resulting in polarization[13, 30, 50]. Due to the lattice distortion, the bond angle of $Cr^{3+}/Fe^{3+}$-$O^{2-}$-$Cr^{3+}/Fe^{3+}$ is slightly less than 180°, causing a deviation of the actual ferromagnetic direction. These tilted magnetic moments cause an internal magnetic field covering the A position, causing the spin direction of $Er^{3+}$ to rotate towards the direction of the internal magnetic field[51]. Since rare earth ions in perovskite structures break space inversion symmetry, $R^{3+}$ ions are located closer to a pair of $Cr^{3+}/Fe^{3+}$ spins that reverse their directions and are subjected to a force along the $Cr^{3+}/Fe^{3+}$-$Er^{3+}$ connecting direction[15]. As shown in **Figure 9**, the net force of $Cr^{3+}/Fe^{3+}$ on Er in a sublattice will cause Er to shift. Due to the difference in the magnetic moment size between $Cr^{3+}$ and $Fe^{3+}$, there is a horizontal component to the displacement of $Er^{3+}$. After forming a stable polarization, $Er^{3+}$ will be coupled with the nearest $Cr^{3+}/Fe^{3+}$ ions, which is relatively complex, with the dominant contribution being the superexchange interaction(3d-2p-4f) between $Er^{3+}$-$O^{2-}$-$Cr^{3+}/Fe^{3+}$[46]. Although the spin of $La^{3+}$ is zero, La doping enhances the spatial asymmetry of the polar lattice, and the 4f orbitals of $La^{3+}$ facilitate strong spin-orbit coupling, both of which contribute to magnetoelectric coupling[52]. In addition, the DMI between $Cr^{3+}/Fe^{3+}$ and $Er^{3+}$ leads to weak magnetization and polarization, but these contributions are not dominant[46].

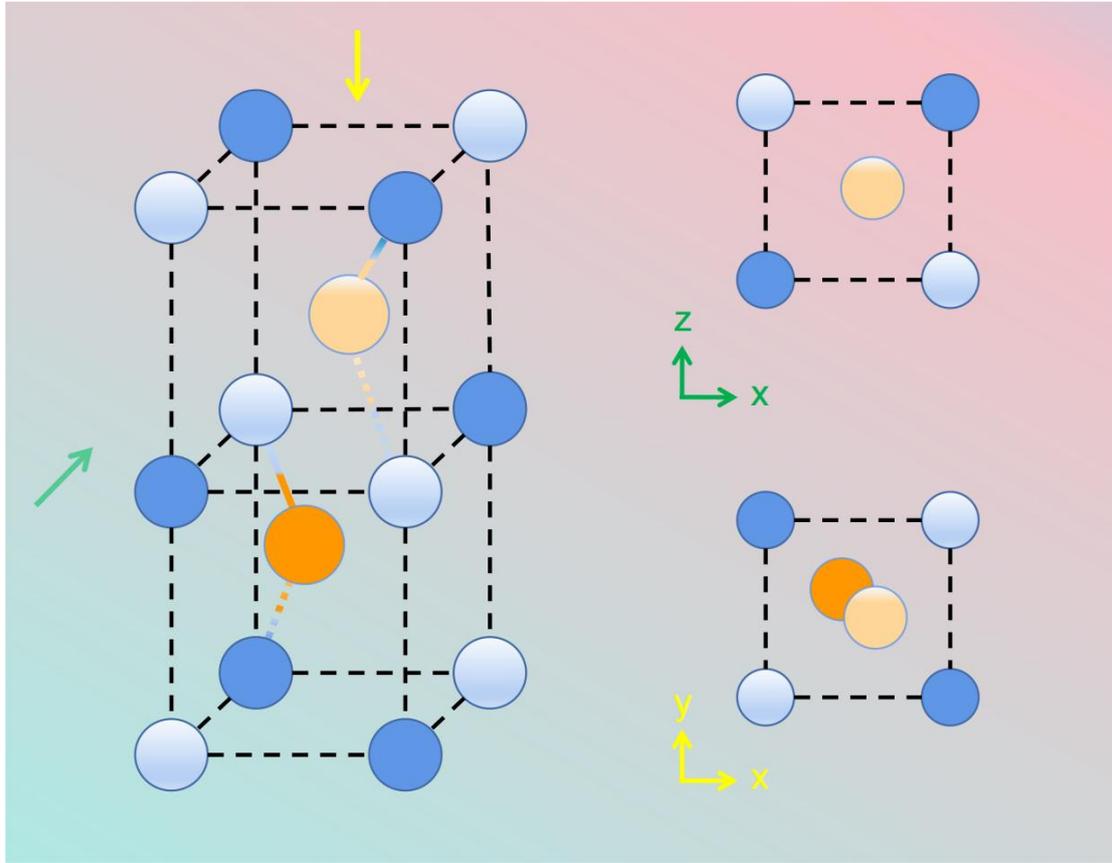

**Fig. 9** Effect of $Cr^{3+}/Fe^{3+}$ on $Er^{3+}$ in ELCFO lattice, where blue and light blue represent $Cr^{3+}/Fe^{3+}$ with opposite spin directions, orange and light orange represent $Er^{3+}$ with opposite spin components in the antiferromagnetic direction, two-colour solid lines indicate attractive forces, and two-colour dashed lines indicate repulsive forces. The two illustrations on the right side clearly show the positions of rare earth ions within the lattice from different directions.

The significant change in the spin directions of $Cr^{3+}$ and $Fe^{3+}$ at $T_{SR'}$ has a significant impact on ELCFO. In addition, as previous studies have shown, the spin direction of $Er^{3+}$ changes between $T_{SR}$ and $T_{SR'}$[53]. This change in spin direction leads to a change in the contribution of $Er^{3+}$ to the multiferroic of ELCFO. As the temperature decreases below $T_{SR}$, the arrangement of $Cr^{3+}/Fe^{3+}$ ions in the ELCFO system shifts to $\Gamma_{12}$, causing a change in the magnetization induced by DMI. At very low temperatures ($T \sim 6.4K$), $Er^{3+}$ undergoes an antiferromagnetic ordering phase transition, leading to a decrease in the magnetization strength of ELCFO, which is reflected in the *dM/dT* vs. *T* curve. The spin arrangement of the G-type antiferromagnetic structure and the change of spin structure in different temperature ranges are shown in **Figure 10**.

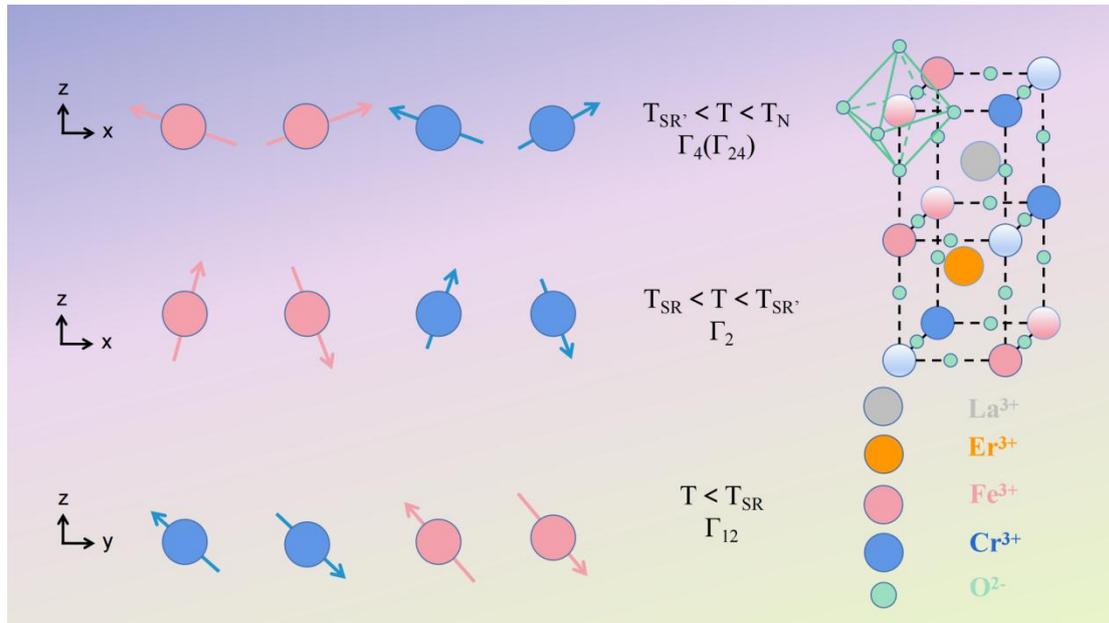

**Fig. 10** The spin direction of ELCFO in different temperature ranges.

## *3.5. Magnetization and Polarization*

The above analysis indicates that the magnetization intensity in the ELCFO system is caused by spin. The ferromagnetic contribution mainly includes weak ferromagnetism induced by DMI between $Cr^{3+}/Fe^{3+}$-$Cr^{3+}/Fe^{3+}$ and $Cr^{3+}/Fe^{3+}$-$Er^{3+}$ that is not canceled out. Due to different magnetic moments, adjacent $Cr^{3+}$ and $Fe^{3+}$ show sub-ferrimagnetism in the weak ferromagnetic perpendicular direction, and the magnetic moment of $Er^{3+}$(as shown in **Figure 11**). Therefore, we can express the magnetization intensity as:

$$M_1 = 1/N \sum_i <S_i^{Cr}> + 1/N \sum_i <S_i^{Fe}> \quad (5)$$

$$M_2 = 1/N \sum_i <S_i^{Er}> + 1/N \sum_i <S_i^{\perp}> \quad (6)$$

$S_i$ represents the Heisenberg spin of $Cr^{3+}$ or $Fe^{3+}$ at site *i*, and *N* represents the number of lattice units in the crystal[54]. **Equation 5** represents the contribution of the magnetic moments of $Cr^{3+}$ and $Fe^{3+}$ to the magnetization strength of the system. In **Equation 6**, the first term represents the contribution of $Er^{3+}$ to the magnetization strength, and the second term represents the magnetization strength generated by the B-site ions perpendicular to the weak ferromagnetic ordering direction.

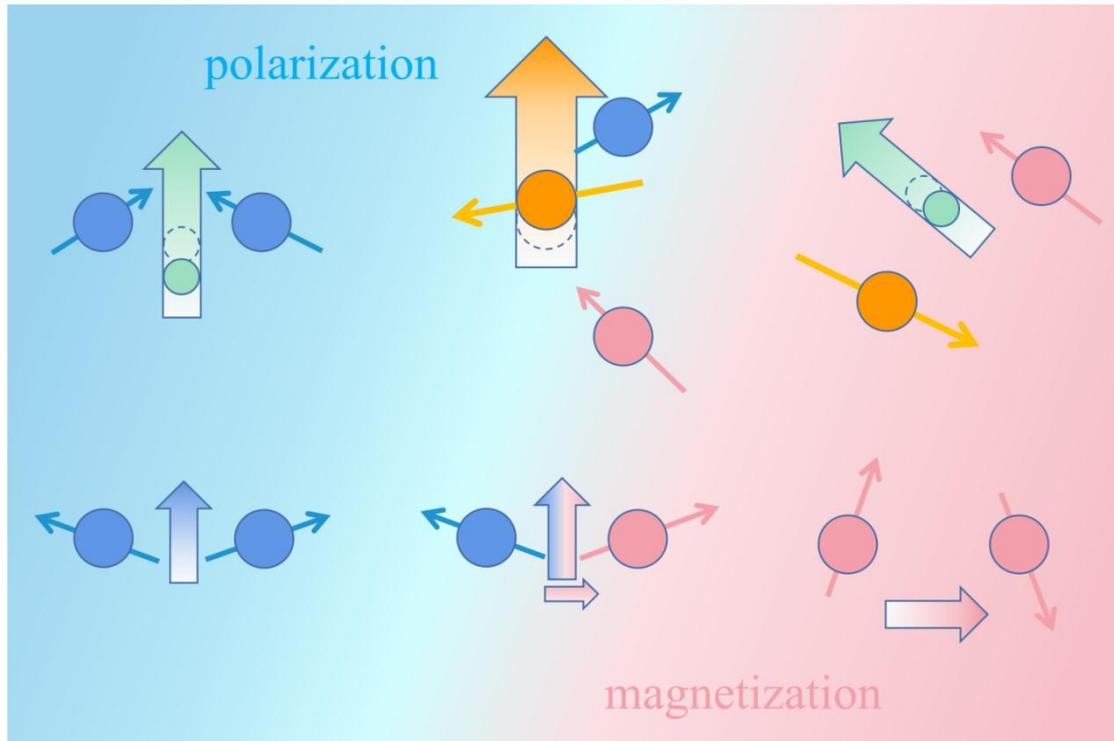

**Fig. 11** As part of the mechanism that produces magnetization and polarization, the thin arrows indicate the ionic spin, and the thick arrows indicate the direction of magnetization or polarization.

The polarization in the ELCFO system arises from the relative displacement of ions, with the dominant source being the polarization resulting from the shift of $Er^{3+}$ along the z-axis, while the rest of the polarization is caused by displacement polarization due to the DMI between the cations (as shown in **Figure 11**). Since the source of polarization is the relative displacement of ions within the lattice, the static displacement is used to express the polarization of ELCFO. By correlating the interaction constants between $Cr^{3+}/Fe^{3+}$ and $Er^{3+}$ and the $R^{3+}$ displacement using the spin-Peierls method, the polarization can be expressed as[46]:

$$P_1 = {e^*\gamma}/{A} \sum_{\beta} \sum_{ij} (e_{ij})^{\beta} < S_i^{Cr/Fe} \cdot S_j^{Er} > \quad (7)$$

$$P_2 = {e^*\lambda_1}/{A} \sum_{\beta} \sum_{ij} < [e_{ij}^{Cr/Fe-Cr/Fe} \times (S_i^{Cr/Fe} \times S_j^{Cr/Fe})]^{\beta} >$$

$$+ {e^*\lambda_2}/{A} \sum_{\beta} \sum_{ij} < [e_{ij}^{Cr/Fe-Er} \times (S_i^{Cr/Fe} \times S_j^{Er})]^{\beta} > \quad (8)$$

The spin-lattice interaction $\lambda$, which results from relativistic spin-orbit interactions, corresponds to DMI when static displacement breaks inversion symmetry. The spin-lattice coupling $\gamma$ is the first-order derivative of the exchange constant between $Cr^{3+}/Fe^{3+}$ and $Er^{3+}$, while $A$ represents the elastic constant, $e^*$ denotes the Bonn effective charge, and $\beta = x, y, z$. The polarization due to $Er^{3+}$ and the neighboring $Cr^{3+}/Fe^{3+}$ can be expressed using **Equation 7**, while the polarization due to DMI can be expressed using **Equation 8**. Magnetic fields will increase the observed polarization, as demonstrated in previous research[18,20].

**Conclusions**

We investigated the effect of ionic spin on the formation of multiferroic by preparing $Er_{0.9}La_{0.1}Cr_{0.8}Fe_{0.2}O_3$ nanopowders using the sol-gel method. XRD tests confirmed conformity to the orthogonal perovskite Pbnm space group. Through tolerance factor $\tau$ and the torsion angle of the octahedron, we can judge that the ELCFO has lattice distortion. The microstress and lattice spacing caused by distortion is obtained from the WH formula. The Mössbauer spectrum at 300 K showed double peaks, indicating that the ELCFO is in a paramagnetic state, and the lattice distortion created an electric field gradient around $Fe^{3+}$. The magnetization curves at room temperature showed a linear correlation, indicating a paramagnetic state of the sample, which corresponds to the Mössbauer spectrum. As the temperature decreased, the magnetization of the magnetization curve increased.

The thermomagnetic curves indicated the complex magnetic characteristics of the system, and we analyzed the states of individual ionic spins and their changes at different temperatures in the ELCFO system. Accordingly, we investigated the origin of multiferroic and the characteristics of magnetization and polarization in this system.By observing the images of magnetization with temperature, we inferred the states of ions, which provides a new way of thinking for the study of materials science.

**Acknowledgements**

This work was supported by National Natural Science Foundation of China (grant number 12105137, 62004143), the Central Government Guided Local Science and Technology Development Special Fund Project (2020ZYYD033), the National Undergraduate Innovation and Entrepreneurship Training Program Support Projects of China, the Natural Science Foundation of Hunan Province, China (grant number S202210555139), the Natural Science Foundation of Hunan Province, China (grant number 2020JJ4517), Research Foundation of Education Bureau of Hunan Province, China (grant number 19A433, 19C1621).

**Conflicts of interest**

There are no conflicts to declare.

**Data Availability Statement**

The datasets generated and/or analysed during the current study are not publicly available due we believe that this can ensure that the experimental data will not be disclosed and help protect our copyright，but are available from the corresponding author on reasonable request.